\documentclass[10pt,conference,a4paper]{IEEEtran}
\usepackage{amsmath,amsfonts}
\usepackage{algorithm}
\usepackage{array}
\usepackage{cite}
\usepackage{verbatim}
\usepackage{stfloats}
\usepackage{tabularx}
\usepackage{lipsum}
\usepackage{url}
\usepackage{graphicx}
\usepackage{tikz}
\usepackage{multirow}
\usepackage{colortbl}
\usepackage{hhline}
\usepackage{rotating}
\usepackage[draft,hidelinks,colorlinks=true,citecolor=blue]{hyperref}

\usepackage{tabulary}
\usepackage{amssymb,amsthm}
\usepackage{tikz}
\usetikzlibrary{calc}
\usepackage{stackengine}
\usepackage{scalerel}
\usepackage{fp}
\usepackage{color,soul}

\usepackage{enumitem}
\usetikzlibrary{svg.path}
\definecolor{orcidlogocol}{HTML}{A6CE39}
\tikzset{orcidlogo/.pic={
\fill[orcidlogocol] svg{M256,128c0,70.7-57.3,128-128,128C57.3,256,0,198.7,0,128C0,57.3,57.3,0,128,0C198.7,0,256,57.3,256,128z};
\fill[white] svg{M86.3,186.2H70.9V79.1h15.4v48.4V186.2z}
                 svg{M108.9,79.1h41.6c39.6,0,57,28.3,57,53.6c0,27.5-21.5,53.6-56.8,53.6h-41.8V79.1z M124.3,172.4h24.5c34.9,0,42.9-26.5,42.9-39.7c0-21.5-13.7-39.7-43.7-39.7h-23.7V172.4z}
                 svg{M88.7,56.8c0,5.5-4.5,10.1-10.1,10.1c-5.6,0-10.1-4.6-10.1-10.1c0-5.6,4.5-10.1,10.1-10.1C84.2,46.7,88.7,51.3,88.7,56.8z};}}
\newcommand\orcidicon[1]{\href{https://orcid.org/#1}{\mbox{\scalerel*{
\begin{tikzpicture}[yscale=-1,transform shape]
\pic{orcidlogo};
\end{tikzpicture}
}{|}}}}
\theoremstyle{definition}

\usepackage{textcomp}

\usepackage{algpseudocode}
\usepackage{subcaption}

\begin{document}
\title{Would Learning Help? Adaptive CRC–QC-LDPC Selection for Integrity in 5G-NR V2X}
\author{
\IEEEauthorblockN{
\makebox[\linewidth][c]{
Sarah Al-Shareeda\IEEEauthorrefmark{2}\IEEEauthorrefmark{4},
Gulcihan Ozdemir\IEEEauthorrefmark{2},
Arouj Fatima\IEEEauthorrefmark{7},
M\u{a}d\u{a}lin-Dorin Pop\IEEEauthorrefmark{11},
Bander A. Jabr\IEEEauthorrefmark{10},
}}
\IEEEauthorblockN{
\makebox[\linewidth][c]{
Yasser Bin Salamah\IEEEauthorrefmark{10}, and
Jacques Demerjian\IEEEauthorrefmark{5}\IEEEauthorrefmark{9}
}}
\IEEEauthorblockA{\IEEEauthorrefmark{2}Informatics Institute, Istanbul Technical University, Turkey}
\IEEEauthorblockA{\IEEEauthorrefmark{4}Center for Automotive Research (CAR), The Ohio State University, USA}
\IEEEauthorblockA{\IEEEauthorrefmark{7}Emerging Energy Contracting Est., Bahrain}
\IEEEauthorblockA{\IEEEauthorrefmark{11}
Computer and Information Technology Department, Politehnica University of Timi\c{s}oara, Romania}
\IEEEauthorblockA{\IEEEauthorrefmark{10}Computer Engineering and Electrical Engineering Departments, King Saud University, Riyadh, Saudi Arabia}
\IEEEauthorblockA{\IEEEauthorrefmark{5}LaRRIS, Faculty of Sciences, Lebanese University, Lebanon}
\IEEEauthorblockA{\IEEEauthorrefmark{9}Faculty of Arts and Sciences, Holy Spirit University of Kaslik (USEK), Lebanon}

\IEEEauthorblockA{
\{alshareeda, ozdemirg\}@itu.edu.tr, 
arouj26@gmail.com,
madalin.pop@upt.ro,\\
\{bjabr,ybinsalamah\}@ksu.edu.sa,
jacques.demerjian@ul.edu.lb
}
}

\markboth{}{}
\maketitle
\begin{abstract}
Vehicle-to-everything (V2X) communications impose stringent physical-layer integrity requirements, particularly under short-packet transmission and mobility-induced channel variation. This paper studies whether standard-compliant online selection of Cyclic Redundancy Check (CRC) polynomials and Quasi-Cyclic Low-Density Parity-Check (QC-LDPC) coding rates can reduce silent (undetected) errors in 5G New Radio (5G-NR) V2X links. The joint configuration problem is formulated as a lightweight Contextual Bandit (CB) with a small, discrete action space, and a discounted LinUCB policy is evaluated against greedy online adaptation and a conservative fixed baseline. A 5G-NR-compliant physical-layer simulation is developed using Sionna, modeling mobility through time-correlated Rayleigh fading, where vehicle speed governs channel correlation, and non-stationary interference via a two-state Markov process. The learning agent operates on coarse receiver feedback, including a noisy Signal-to-Noise Ratio (SNR) estimate and indicators of burst interference and deep fades, and targets minimization of the Undetected Error Probability (\(P_{UE}\)) while accounting for the Detected Error Probability (\(P_{DE}\)). Overall, our objective is to delineate the mobility regimes in which learning-assisted CRC–QC-LDPC configuration improves physical-layer integrity in 5G-NR V2X systems. Our results indicate that learning-assisted adaptation is most effective at low to moderate mobility, reducing \(P_{UE}\) by up to 50–70\% relative to greedy selection in the low-SNR regime (\(-5\) to 5~dB) and approaching the best fixed configuration at higher \(E_b/N_0\). At high mobility (\(\geq 180\)~km/h), fast channel decorrelation weakens temporal predictability, limiting the effectiveness of online learning and reducing performance differences across policies.
\end{abstract}

\IEEEpeerreviewmaketitle
\section{Introduction}\label{intro}
\begin{figure}[!htbp]
\centering
\includegraphics[width=.8\columnwidth]{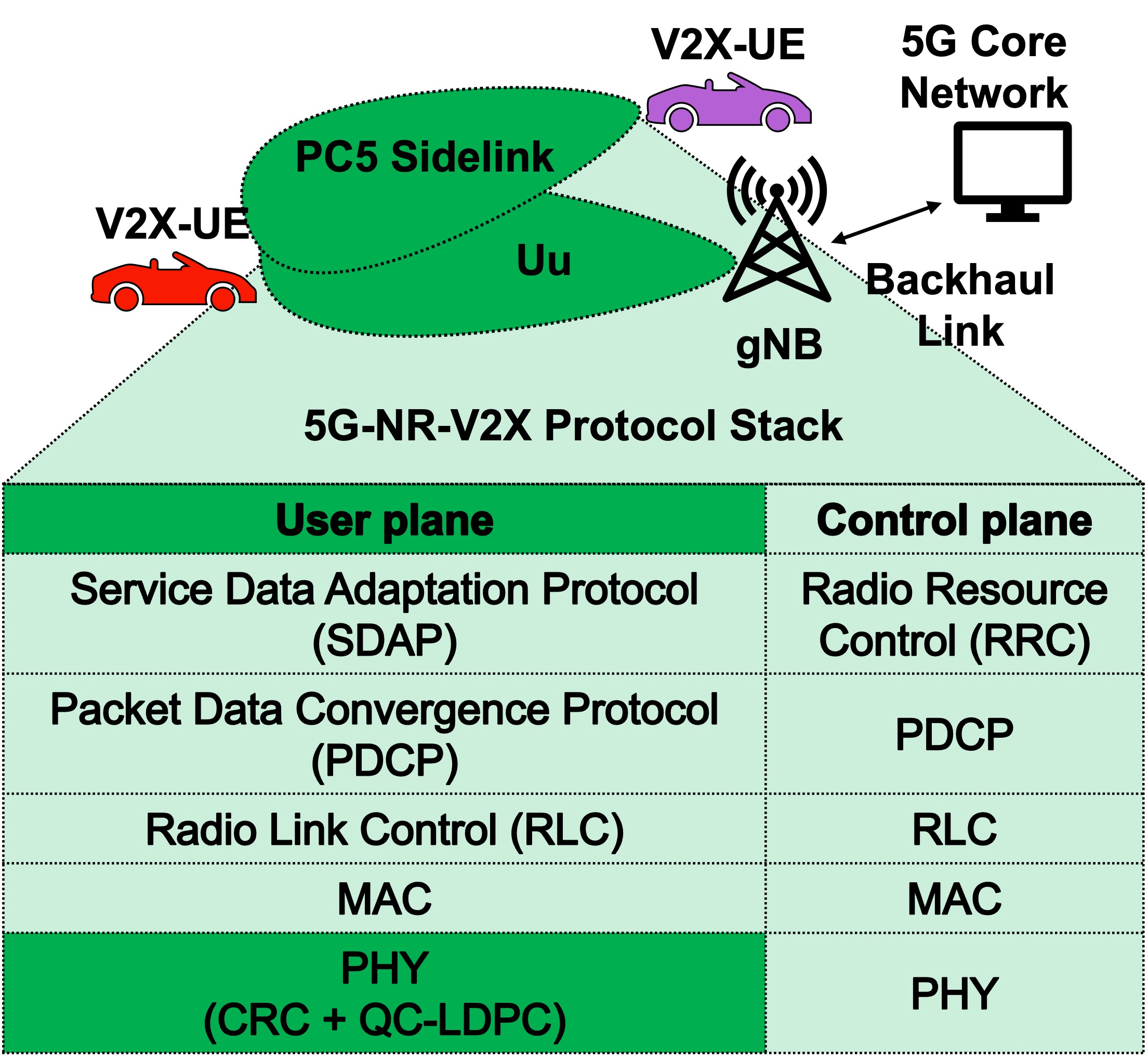}
\caption{\textcolor{black}{5G-NR V2X protocol stack supporting both the direct PC5 sidelink and the Uu air interface, highlighting PHY-layer CRC-based error detection and QC-LDPC-based error correction mechanisms considered in this work.}}
\label{fig:arc}
\end{figure}

5G New Radio (5G-NR) systems are designed to support Vehicle-to-Everything (V2X) services operating under Ultra-Reliable Low-Latency Communication (URLLC) requirements, where stringent reliability targets and millisecond-level latency constraints are fundamental to safety-critical operation \cite{5g-nr}. At the physical layer, see Fig. \ref{fig:arc}, integrity is ensured through a concatenated error-control architecture that combines Cyclic Redundancy Check (CRC) codes for error detection with powerful Forward Error Correction (FEC) schemes, most notably Quasi-Cyclic Low-Density Parity-Check (QC-LDPC) codes. These mechanisms are standardized and highly optimized; however, their configuration remains largely rule-based and quasi-static, determined primarily by payload length, service type, and nominal coding rate, rather than instantaneous channel dynamics or decoder outcomes. Such fixed, standard-compliant configurations are designed through extensive offline analysis and have proven effective under quasi-stationary channel assumptions, prioritizing determinism, low complexity, and certification feasibility \cite{morais20255g}.

In highly mobile V2X environments, wireless channel conditions can vary rapidly due to mobility-induced fading, Doppler effects, and bursty interference \cite{11054162,al2017enhancing}. These dynamics can rapidly invalidate the assumptions under which offline physical-layer configurations are selected. While fixed configurations are robust under nominal operating conditions, their lack of adaptability may lead to inconsistent performance under such dynamics, particularly in short-packet integrity-sensitive regimes. In these regimes, Undetected Errors (UE) are of particular concern, as erroneous packets that pass CRC verification cannot be mitigated through retransmission and may silently corrupt higher-layer processing, directly compromising physical-layer integrity \cite{8904104} where even small increases in the UE Probability ($P_{UE}$) can have a disproportionate impact on safety-critical operation. Recent performance evaluations of 5G-NR V2X sidelink communication further emphasize that maintaining low latency and high reliability under realistic vehicular mobility remains challenging, especially in the presence of dynamic interference and fast channel decorrelation \cite{11086589}.

\begin{figure}[!htbp]
\centering
\includegraphics[width=\columnwidth]{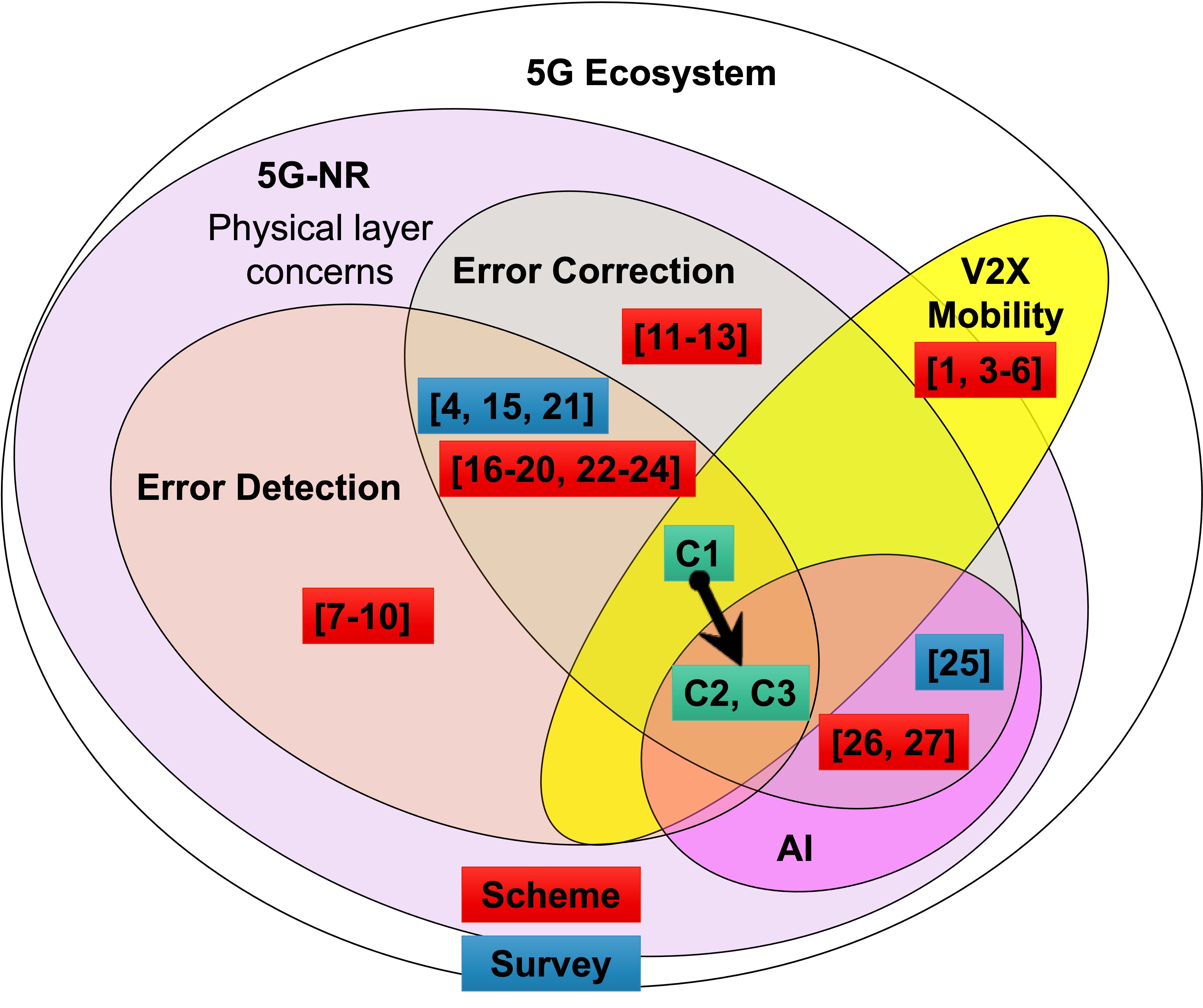}
\caption{\textcolor{black}{Literature landscape and Addressed Research Question.}}
\label{fig:contributions}
\end{figure}

A substantial body of literature has investigated error-control design and optimization in 5G-NR systems, as depicted in Fig. \ref{fig:contributions}. Prior studies examine CRC polynomial design, decoder optimizations, and integrated coding architectures \cite{koopman,abdel2025encoding,baicheva2024some,naji2025netcrc,ren2024generalized,verma2024high,lin2025hybrid}, as well as joint CRC–LDPC and CRC–polar coding strategies for reducing the $P_{UE}$ \cite{alnajjar2024channel,sy2024demystifying,egilmez2022soft,zhu2022adaptive,zhan2024high,sauter2025undetected,sinha2024performance,iqbal2023hardware,belhadj2021error,wang2023probabilistic,azeem2022exploiting}. In parallel, Machine Learning (ML) techniques have been proposed for decoder-level adaptation and parameter tuning \cite{indoonundon2023ai,hernandez20255g,gunturu2021machine}. While these approaches demonstrate performance gains under static or mildly varying conditions, they predominantly assume offline configuration, decoder-centric adaptation, or stationary channel behavior. Consequently, no existing studies address online, learning-assisted joint CRC and QC-LDPC configuration under mobility-induced non-stationarity using standard-compliant components, nor do they explicitly analyze regimes in which learning-based adaptation may adversely affect physical-layer integrity. This work addresses this challenge by proposing a learning-assisted framework for adaptive selection of CRC and QC-LDPC configurations in 5G-NR V2X systems. The joint selection of error detection and error correction parameters is formulated as a lightweight online learning problem using a Contextual Bandit (CB) framework. The CB learning agent operates on coarse channel context, including a noisy Signal-to-Noise Ratio (SNR) estimate and indicators of bursty interference and deep-fade events, and selects, at each transmission opportunity, a CRC polynomial and a QC-LDPC coding rate to minimize $P_{UE}$ while accounting for the Detected Error Probability ($P_{DE}$). The main contributions of this paper are summarized as follows:
\begin{enumerate}[label=C\arabic*.]
\item we simulate a 5G-NR-compliant physical-layer framework incorporating 3GPP CRC polynomials, true 5G-NR QC-LDPC construction, Binary Phase-Shift Keying (BPSK) transmission, mobility-induced time-correlated Rayleigh fading, bursty interference, soft-input Belief-Propagation (BP) decoding, and CRC verification.
\item we formulate a joint CRC–QC-LDPC configuration problem as a CB task, enabling efficient online adaptation over a small, discrete, standard-compliant decision space. The proposed CB formulation aligns naturally with the discrete and standardized configuration space of 5G-NR, enabling fast online adaptation without modifying existing procedures or introducing significant computational overhead.
\item we design of an integrity-aware learning objective centered on minimizing the $P_{UE}$ while accounting for $P_{DE}$, enabling systematic identification of operating regimes in which learning improves integrity robustness and those in which fixed configurations remain preferable.
\end{enumerate}

Through comprehensive evaluation across CRC types, QC-LDPC code rates, and mobility levels, this study identifies regimes in which learning-assisted adaptation improves integrity robustness and those in which mobility-induced non-stationarity renders fixed configurations more reliable than adaptive policies. Specifically, high mobility can invalidate the temporal consistency assumptions required for effective learning, making fixed conservative configurations preferable in some URLLC V2X scenarios. The remainder of the paper is organized as follows. Sections~\ref{sec:system_model} presents the proposed learning-assisted joint configuration framework. Section~\ref{sec:results} reports experimental results and analysis. Finally, Section~\ref{conc} concludes the work and discusses directions for future work.

\section{System Model and Learning-Assisted Configuration Framework}
\label{sec:system_model}
\begin{figure*}[!htbp]
\centering
\includegraphics[width=\textwidth]{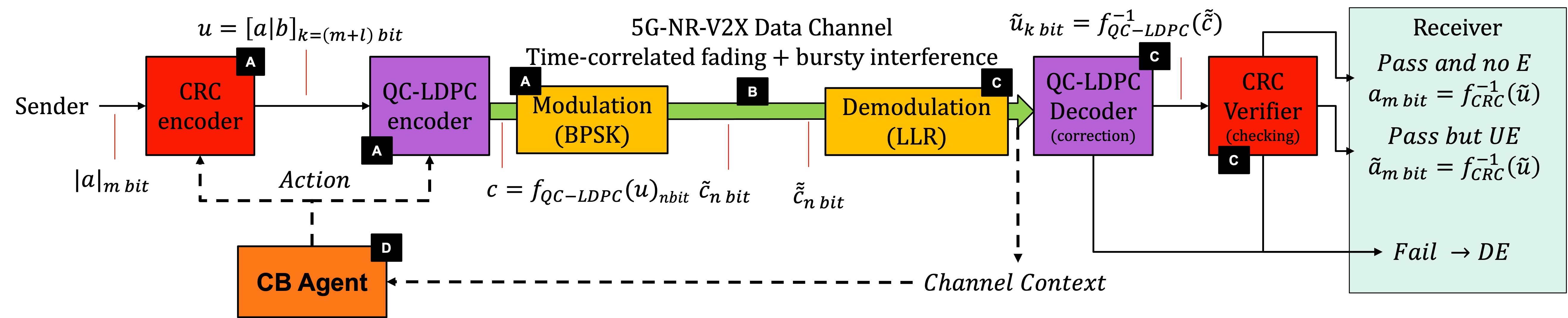}
\caption{Our Learning-Assisted standard-compliant 5G-NR physical-layer processing chain.}
\label{fig:chain}
\end{figure*}

This section formalizes our system model, depicted in Fig. \ref{fig:chain}, and the proposed learning-assisted framework for adaptive joint selection of CRC polynomials and QC-LDPC coding rates in 5G-NR V2X links. We consider a single-link 5G-NR V2X physical-layer transmission between one transmitter and one receiver. Medium-access contention, packet collisions, sidelink resource competition, and multi-user scheduling effects are intentionally excluded in order to isolate physical-layer integrity effects arising from coding, decoding, and mobility-induced channel dynamics. The model captures how a standard 5G-NR link behaves under mobility and limited channel knowledge, and uses this to frame adaptive CRC–QC-LDPC selection as a lightweight CB problem.

\subsection{Transmitter-Side Physical-Layer Encoding Model}
We consider packet-based 5G-NR physical-layer transmissions over discrete time instances indexed by $t$. At each instance, an information block
\begin{equation}
a_t \in \{0,1\}^{m}
\end{equation}
is processed by a standard-compliant CRC-QC-LDPC encoding chain. CRC encoding interprets $a_t$ as a polynomial over ${GF}(2)$ and appends $l$ parity bits generated by dividing $a_t(x)x^l$ by a generator polynomial $g_{\mathrm{CRC}}(x)$ of degree $l$. The resulting remainder $b_t$ is concatenated with the payload $a_t$ to form a CRC-augmented sequence
\begin{equation}
u_t \in \{0,1\}^{k}, \quad k = m + l.
\end{equation}

In this work, we consider standard CRC polynomials defined by 3GPP TS~38.212, namely CRC-6, CRC-11, CRC-16, and CRC-24A, which offer different trade-offs between detection capability and redundancy overhead and are routinely employed in 5G-NR systems. In short-packet V2X transmissions, CRC effectiveness is particularly critical, as limited redundancy and structured decoding errors can increase the likelihood of UEs. The CRC-protected sequence $u_t$ is subsequently encoded using a 5G-NR QC-LDPC encoder constructed from standardized Base Graphs (BG~1 or BG~2). A lifting factor $Z$ expands the selected BG into a full parity-check matrix; the encoded codeword is generated via the corresponding generator matrix $G$ as
\begin{equation}
c_t = u_t G \bmod 2\,\in \{0,1\}^{n}.
\end{equation}
Due to their structured nature, QC-LDPC codes exhibit non-random residual decoding errors under soft-decision BP decoding, which interact non-trivially with CRC detection. Consequently, CRC and LDPC performance cannot be treated independently when assessing $P_{UE}$. Next, $c_t$ is mapped to BPSK symbols according to $0 \mapsto +1$ and $1 \mapsto -1$, yielding $\tilde{c}_t \in \{-1,+1\}^n$ codeword to be transmitted over the channel described below.

\subsection{Mobility-Affected Channel and Interference Model}
Transmission of $\tilde{c}_t$ takes place in complex baseband over a flat-fading wireless channel with additive noise,
\begin{equation}
\tilde{\tilde{c}}_t = h_t \tilde{c}_t + w_t,
\end{equation}
where $h_t \in \mathbb{C}$ denotes the complex channel coefficient capturing amplitude fading and phase rotation, and $w_t$ represents circularly symmetric complex Additive White Gaussian Noise (AWGN). To capture V2X mobility effects, the channel coefficient $h_t$ evolves according to a time-correlated Rayleigh fading process,
\begin{equation}
h_t = \rho h_{t-1} + \sqrt{1-\rho^2}\, z_t,
\end{equation}
where $z_t \sim \mathcal{CN}(0,1)$ is an i.i.d. complex Gaussian innovation term and $\rho \in [0,1]$ controls the temporal correlation of the channel. The correlation coefficient $\rho$ depends on the relative transmitter-receiver speed through the maximum Doppler frequency
\begin{equation}
f_D = \frac{v}{s} f_0,
\end{equation}
where $v$ denotes the relative vehicle speed, $f_0$ is the carrier frequency, and $s$ is the speed of light. Under the classical Jakes model, the one-step temporal correlation over the transmission interval $T$ is approximated as
\begin{equation}
\rho \approx J_0\!\left(2\pi f_D T\right),
\end{equation}
with $J_0(\cdot)$ denoting the zero-order Bessel function of the first kind. Beyond fading and thermal noise $w_t$, the received signal may also be impaired by intermittent non-stationary external interference, which is modeled explicitly using a two-state Markov process. Let $I_t \in \{0,1\}$ denote the interference state at transmission instance $t$, where $I_t = 1$ corresponds to a burst-interference state. The interference dynamics follow
\begin{equation}
\Pr(I_t = 1 \mid I_{t-1} = 0) = p_{01}, \quad
\Pr(I_t = 1 \mid I_{t-1} = 1) = p_{11}.
\end{equation}
When interference is active, the effective noise variance experienced at the receiver increases accordingly. The interpretation and normalization of the noise term $w_t$ w.r.t. the operating SNR per bit $E_b/N_0$ are specified in the receiver-side processing described in the following subsection.

\subsection{Receiver-Side Physical-Layer Decoding}\label{decode}
At the receiver, the noisy channel output $\tilde {\tilde c}$ is processed to recover the transmitted information and to assess physical-layer integrity. The operating signal quality is parameterized by $E_b/N_0$. For coded transmission with effective rate $R_t$, the corresponding symbol-level SNR under BPSK modulation satisfies $R_t\,E_b/N_0$. In the complex baseband model, thermal noise $w_t$ is represented as circularly symmetric complex AWGN with variance $\sigma^2$ per complex dimension. The receiver employs a nominal noise variance
\begin{equation}
\sigma^2 = \frac{1}{2 R_t \cdot 10^{\frac{E_b/N_0}{10}}},
\end{equation}
which is consistent with the assumed BPSK signaling and is used for soft demodulation unless otherwise stated.

At the receiver, coherent soft demodulation is performed using Log-Likelihood Ratios (LLRs). For the $i^{\text{th}}$ received symbol at transmission instance $t$, the LLR is computed as
\begin{equation}
L_{t,i} = \frac{2\,\Re\{h_t^* \tilde{\tilde{c}}_{t,i}\}}{\sigma^2}.
\end{equation}
The LLRs quantify the reliability of the received symbols and serve as soft inputs to an iterative BP decoder operating on the Tanner graph of the selected QC-LDPC code. If the BP decoder fails to converge within the prescribed number of iterations, the decoding attempt is immediately classified as a DE, and CRC verification is not performed. If the decoder converges, it outputs an estimate $\tilde{u}_t$ of the CRC-protected systematic bit sequence $u_t$, which is subsequently subjected to CRC verification using the same CRC polynomial applied at the transmitter. If the CRC check fails, the transmission is classified as a DE.

An UE occurs when $\tilde{u}_t \neq u_t$ while CRC verification succeeds, causing the erroneous payload to be accepted as valid. Over a horizon of $M$ transmissions, the $P_{UE}$ is defined as
\begin{equation}\label{eq:pue}
P_{UE} = \frac{1}{M} \sum_{t=1}^{M} UE_t.
\end{equation}
The $P_{UE}$ is therefore a critical integrity metric in reliability- and safety-oriented communication systems. Unlike DEs, UEs cannot be mitigated through retransmission or higher-layer recovery mechanisms once accepted, resulting in silent data corruption. This risk is particularly severe in short-packet V2X communications under high mobility, where channel non-stationarity and structured residual decoding errors increase the likelihood of CRC verification failure. Accordingly, minimizing $P_{UE}$ constitutes the primary physical-layer integrity objective of this work.

\subsection{Learning-based CRC-LDPC Configuration Framework}
Having defined the transmitter processing, mobility-affected channel, and receiver-side integrity events, we now address the core objective of this work: adaptive selection of CRC-QC-LDPC configurations to minimize \eqref{eq:pue} under time-varying V2X conditions. To achieve this objective, we employ a lightweight learning agent based on a
CB formulation. At each transmission instance, the CB observes a low-dimensional context derived from delayed receiver-side feedback, selects a joint CRC-QC-LDPC configuration, and receives a scalar reward reflecting both throughput and physical-layer integrity. This interaction pattern naturally maps the adaptive configuration problem to a contextual multi-armed bandit setting, where learning is driven by observed transmission outcomes rather than explicit channel state information.

\subsubsection{Context Definition under Partial Observability}
The transmitter has no access to the instantaneous channel realization $h_t$ or the instantaneous interference state $I_t$. Instead, it relies on a low-dimensional context vector constructed from delayed and imperfect receiver-side feedback,
\begin{equation}
observation_t =
\begin{bmatrix}
1 &
\lambda_{t-1} &
\lambda_{t-1}^2 &
I_{t-1} &
\mathbb{I}_{{deep},t-1}
\end{bmatrix}^{\mathsf{T}} .
\end{equation}
Here, $\lambda_{t-1}$ denotes a noisy estimate (in dB) of the effective received SNR during the previous transmission instance. This estimate captures the combined effects of the channel realization $h_{t-1}$, the selected coding rate $R_{t-1}$, and the operating $E_b/N_0$, and is consistent with the signal and noise models defined earlier. The quadratic term $\lambda_{t-1}^2$ is included to account for mild nonlinear dependence of decoding reliability on the observed signal quality within an otherwise linear contextual model. The indicator $I_{t-1}$ corresponds to the interference state in the two-state Markov model defined earlier, with $I_{t-1}=1$ indicating the presence of burst interference during the previous transmission. The term $\mathbb{I}_{{deep},t-1}$ denotes the occurrence of a deep fade, inferred at the receiver when the channel magnitude $|h_{t-1}|$ falls below a predefined threshold. Both indicators are derived from receiver-side observations and fed back in a highly quantized form.

\subsubsection{Configuration Space and Action Definition}
At each transmission instance $t$, the transmitter selects a joint CRC–QC-LDPC configuration
\begin{equation}
action_t^{(\mathrm{cfg})} \in \mathcal{A},
\end{equation}
from a finite, standard-compliant configuration set $\mathcal{A} = \{(\mathrm{CRC}_i, R_j)\}$, where $\mathrm{CRC}_i$ denotes a selected CRC polynomial and $R_j$ denotes a valid 5G-NR
QC-LDPC coding rate. The block length $n$ is fixed, while the effective coding rate
\begin{equation}
R_t = \frac{k}{n}
\end{equation}
varies with the selected CRC length and QC-LDPC rate-matching configuration. In this study, four CRC polynomials and three LDPC rates are considered, yielding a total of $|\mathcal{A}| = 12$ possible configurations.

\subsubsection{Integrity-Aware Reward Model}
Each transmission instance yields a scalar reward that prioritizes physical-layer integrity while incorporating throughput. The reward at transmission instance $t$ is defined as
\begin{equation}
reward_t =
R_t \, \mathbb{I}\{DE_t = 0 \land UE_t = 0\}
- \Omega_{{DE}}\, DE_t
- \Omega_{{UE}}\, UE_t,
\end{equation}
where $DE_t \in \{0,1\}$ and $UE_t \in \{0,1\}$ denote the DE and UE indicators defined in \ref{decode}. The weights satisfy $\Omega_{{UE}} > \Omega_{{DE}} > 0$, reflecting the more severe impact of UEs on data integrity.

Having formulated adaptive CRC–QC-LDPC selection as a CB problem, a concrete learning algorithm is required to map observed context to configuration decisions while balancing exploration and exploitation under non-stationary channel conditions. Among CB methods, Linear Upper Confidence Bound (LinUCB) algorithms are particularly well suited to this setting due to their ability to exploit a linear reward structure, operate with low computational complexity, and provide principled uncertainty-driven exploration. To further account for mobility-induced non-stationarity and intermittent interference, a discounted variant of LinUCB is adopted, which gradually downweights outdated observations and enables rapid adaptation to evolving wireless conditions.

At each transmission instance $t$, the transmitter observes the context vector $observation_t$ and selects a configuration according to
\begin{equation}
\begin{aligned}
action_t^{(\mathrm{cfg})} =&
\arg\max_{action \in \mathcal{A}}
\Big(
{\theta}_{action}^{\mathsf{T}}\, observation_t
+\\
&\alpha
\sqrt{
observation_t^{\mathsf{T}}
A_{action}^{-1}
observation_t
}
\Big),
\end{aligned}
\end{equation}
where $\alpha > 0$ controls the exploration–exploitation trade-off. Here, ${\theta}_{action}$ denotes the current estimate of the linear reward model associated with configuration $action$, and $A_{action}$ is a regularized covariance matrix capturing the discounted second-order statistics of past context observations for that configuration.

After the transmission outcome is observed, only the statistics associated with the selected configuration $action_t^{(\mathrm{cfg})}$ are updated using an exponential discount factor $\gamma \in (0,1]$:
\begin{equation}\label{eq:a}
A_{action_t} \leftarrow \gamma A_{action_t}
              + observation_t\, observation_t^{\mathsf{T}},
\end{equation}
\begin{equation}\label{eq:b}
B_{action_t} \leftarrow \gamma B_{action_t}
              + reward_t\, observation_t,
\end{equation}
where $B_{action}$ accumulates the discounted correlation between observed rewards and context features for configuration $action$. The corresponding parameter estimate is then obtained as
\begin{equation}\label{eq:theta}
{\theta}_{action} = A_{action}^{-1} B_{action}.
\end{equation}

The complete learning-assisted CRC–QC-LDPC configuration procedure is summarized in Algorithm~\ref{alg:linucb}.

\begin{algorithm}[!hbp]
\caption{Discounted LinUCB for Adaptive CRC-QC-LDPC Selection}
\label{alg:linucb}
\begin{algorithmic}[1]
\State \textbf{Initialize:}
$A_{action} \gets I$, $B_{action} \gets {0}$ for all $action \in \mathcal{A}$
\For{each transmission instance $t = 1,2,\dots$}
    \State Observe context vector $observation_t$
    \For{each $action \in \mathcal{A}$}
        \State Compute parameter estimate ${\theta}_{action}$ \eqref{eq:theta}
        \State Compute
        \[
        \begin{aligned}
        UCB_{action} =
         {\theta}_{action}^{\mathsf{T}}\, observation_t \\
         + \alpha \sqrt{
            observation_t^{\mathsf{T}}
            A_{action}^{-1}
            observation_t
        }
        \end{aligned}
        \]
    \EndFor
    \State Select
    $action_t^{(\mathrm{cfg})} = \arg\max_{action \in \mathcal{A}} UCB_{action}$
    \State Transmit using the selected CRC-QC-LDPC pair
    \State Observe $reward_t$
    \State Update statistics for $action_t^{(\mathrm{cfg})}$ as in \eqref{eq:a} and \eqref{eq:b}
\EndFor
\end{algorithmic}
\end{algorithm}

This framework enables efficient online adaptation while explicitly exposing the sensitivity of learning-based configuration to mobility-induced non-stationarity, which is examined in the subsequent performance evaluation.

\section{Simulation and Results Discussion}\label{sec:results}
This section evaluates whether {standard-compliant} online selection of CRC and QC-LDPC configurations can reduce the $P_{UE}$ under mobility-induced channel non-stationarity, and identifies the mobility regimes in which learning-based adaptation remains beneficial vs. those in which conservative fixed configurations are preferable. The evaluation is intentionally restricted to a single transmitter-receiver link in order to isolate physical-layer integrity effects arising from coding, decoding, and time-varying channel dynamics. Medium-access contention, packet collisions, sidelink resource competition, and multi-user scheduling are excluded and left for future work. All experiments are implemented in Python~3.11 using the Sionna physical-layer library to ensure faithful realization of 5G-NR QC-LDPC construction, soft-input BP decoding, and CRC verification. Simulations are executed on Google Colab using NVIDIA A100 GPUs \cite{10454851,11106806}. Hardware acceleration is used solely to enable large-scale rare-event Monte Carlo evaluation and does not affect the underlying physical-layer models or learning behavior. Table~\ref{tab:sim_params} summarizes the simulation parameters and environment settings.

\begin{table}[!htbp]
\caption{Simulation Parameters and Settings}
\label{tab:sim_params}
\centering
\begin{tabular}{m{2.9cm} m{5cm}}
\hline
\textbf{Parameter} & \textbf{Value / Description} \\
\hline
Payload length $m$ & $256$ information bits per packet \\
CRC candidates $\mathcal{C}$ &
CRC-6 ($0x27$), CRC-11 ($0x307$), CRC-16 ($0x1021$), CRC-24A ($0x1864CFB$) \\
LDPC block length $n$ & $576$ \\
LDPC information size $k$ & $\{288,384,432\}$ \\
Code rates $R$ & $\{1/2,\,2/3,\,3/4\}$ \\
BPSK Modulation & $0\mapsto +1$, $1\mapsto -1$ \\
Carrier frequency $f_c$ & $5.9$~GHz \\
Transmission interval $T$& $1$~msec \\
Mobility regimes $v$& $\{0,60,120,180,250\}$~km/h \\
SNR observation noise & $\mathcal{N}(0,0.8^2)$~dB added to $\hat{\gamma}$ \\
Learning algorithm & Discounted LinUCB: $\alpha=2.0$, $\gamma=0.998$, $\lambda=1.0$ \\
Training length & $T_{\text{train}}=6000$ steps per speed (LinUCB and Greedy) \\
Validation length & $T_{\text{val}}=6000$ steps per speed (FixedBest selection) \\
Evaluation $E_b/N_0$ grid & $\{-5,0,5,10,15,20,25\}$~dB \\
Rare-event trials & $30{,}000$ trials per $(v,E_b/N_0,\text{policy})$ point \\
URLLC latency model & $B=20$~MHz, TTI=$1$~msec, base stack $0.5$~msec, iter cost $0.08$~msec \\
Packet delay budget & $\text{PDB}=5$~msec \\
Reward weights & $w_g=1.0$, $w_{DE}=0.2$, $w_{UE}=5.0$, $w_{DM}=3.0$ \\
\hline
\end{tabular}
\end{table}

\subsection{Integrity Metrics and Rare-Event Evaluation}
The primary integrity metric is the $P_{UE} = \Pr\{\text{CRC passes} \land \tilde{a}_t \neq a_t\}$, which captures silent corruption events that cannot be mitigated by retransmission or higher-layer recovery once a CRC check passes. For completeness, the $P_{DE}$ is also reported. Both probabilities are estimated empirically over a finite number of transmission trials. Because UEs are rare at moderate-to-high $E_b/N_0$, evaluation is conducted in a rare-event regime. When no UEs are observed, a conservative 95\% confidence upper bound is reported.

\subsection{Policies Compared: Online Adaptation vs Conservative Fixing}\label{subsec:policies}
We compare three policies that represent distinct adaptation philosophies:
\begin{enumerate}
    \item \textbf{Discounted LinUCB (proposed):} an uncertainty-aware CB that balances exploitation and exploration using confidence bounds, with exponential discounting to prioritize recent observations and track non-stationarity.
    \item \textbf{Greedy contextual policy:} a purely myopic online strategy that selects the configuration with the best empirical performance so far (equivalently LinUCB with $\alpha=0$), and therefore reacts strongly to noisy feedback and may oscillate under non-stationarity.
    \item \textbf{FixedBest:} an offline-selected single configuration that minimizes $P_{UE}$ for a given mobility regime and then remains fixed during operation, serving as a conservative integrity benchmark.
\end{enumerate}
This comparison directly tests whether online adaptation remains beneficial as mobility increases, or whether a fixed conservative configuration provides stronger integrity guarantees under severe non-stationarity.

\subsection{Online Adaptation Dynamics Under Mobility}
\begin{figure}[!htbp]
\centering
\includegraphics[width=\columnwidth]{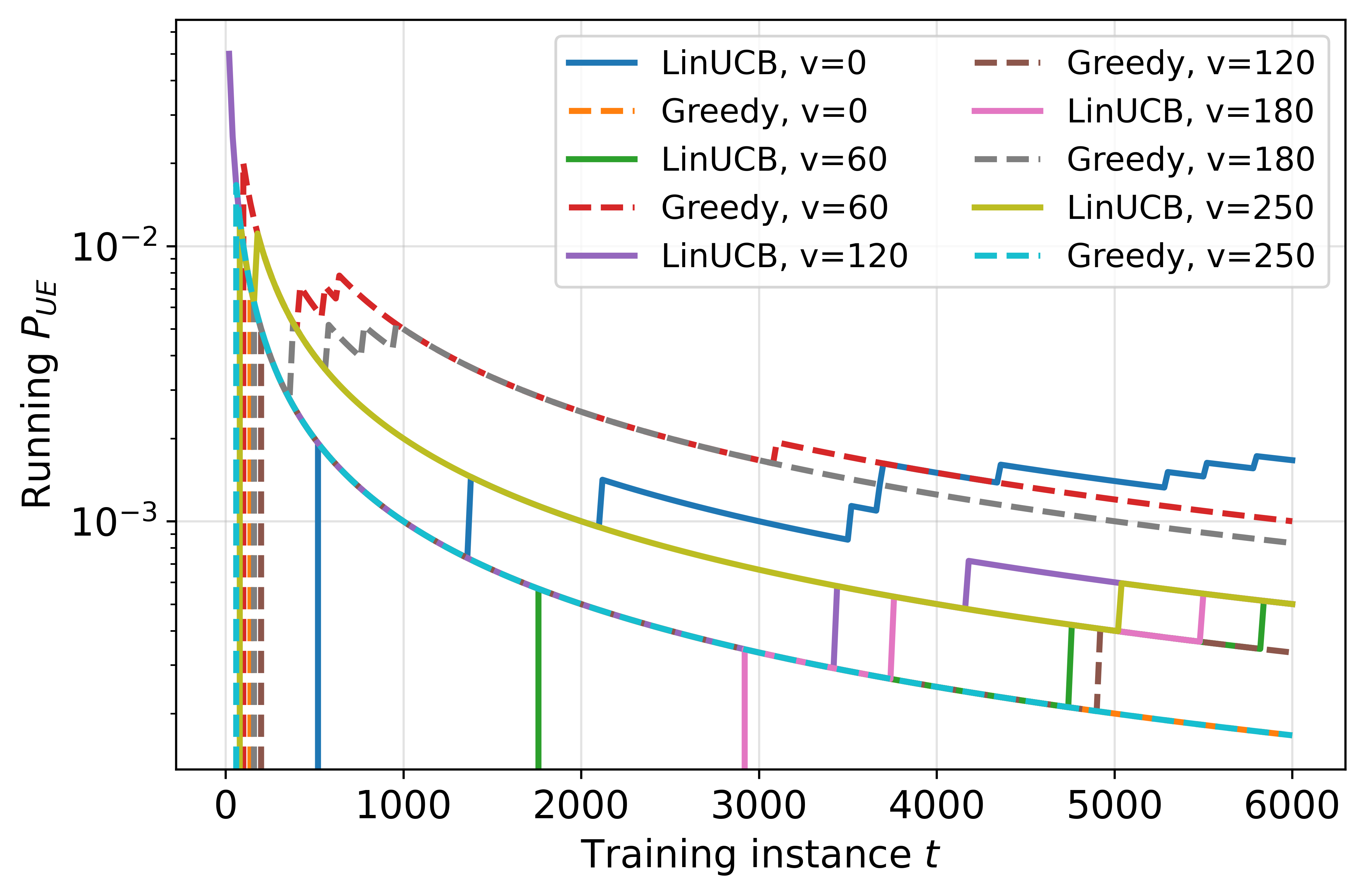}
\caption{Online learning dynamics under mobility-induced channel variation.}
\label{fig:running_pue_all}
\end{figure}

Fig.~\ref{fig:running_pue_all} reports the running-average $P_{UE}$ during training for LinUCB and greedy policies across mobility regimes. At $v=0$~km/h, both policies converge rapidly and stably, consistent with near-stationary channel conditions. As mobility increases, convergence slows and variability increases due to reduced temporal correlation in the fading process and noisier context-feedback alignment. At high mobility ($v=180$ and $250$~km/h), both policies exhibit persistent fluctuations, indicating that short-term observations become less predictive of near-future conditions. This behavior anticipates the regime in which learning-based adaptation may lose its advantage.

\subsection{Decoder Quality Check via $P_{DE}$}
Before interpreting UE trends, we verify that the underlying BP decoding behavior is not itself changing across policies in a way that could confound integrity comparisons. As defined in Section~\ref{decode}, DEs arise from BP decoder failure or CRC rejection and therefore reflect intrinsic decoder reliability under the operating channel conditions. Table~\ref{tab:pde_summary} summarizes $P_{DE}$ across all policies, mobility regimes, and $E_b/N_0$ values. The mean $P_{DE}$ remains tightly bounded (24\%-26\%) with differences below 1.3\% points, and no systematic dependence on policy or mobility is observed. This indicates that the observed integrity differences if any are primarily driven by the CRC-QC-LDPC {configuration choices} rather than by policy-dependent decoder behavior.

\begin{table}[!htbp]
\caption{Decoder Quality via $P_{DE}$}\label{tab:pde_summary}
\centering
\begin{tabular}{lccc}
\hline
Policy & Mean & Max & Min \\
\hline
LinUCB     & 0.2525 & 0.9996 & $<10^{-6}$ \\
Greedy     & 0.2449 & 0.9098 & $<10^{-6}$ \\
FixedBest  & 0.2574 & 0.9789 & $1.67\times10^{-4}$ \\
\hline
\end{tabular}
\end{table}

\subsection{$P_{UE}$ vs. $E_b/N_0$ Across Mobility Regimes}\label{subsec:pue_ebn0}
\begin{figure*}
    \centering
    \includegraphics[width=\linewidth]{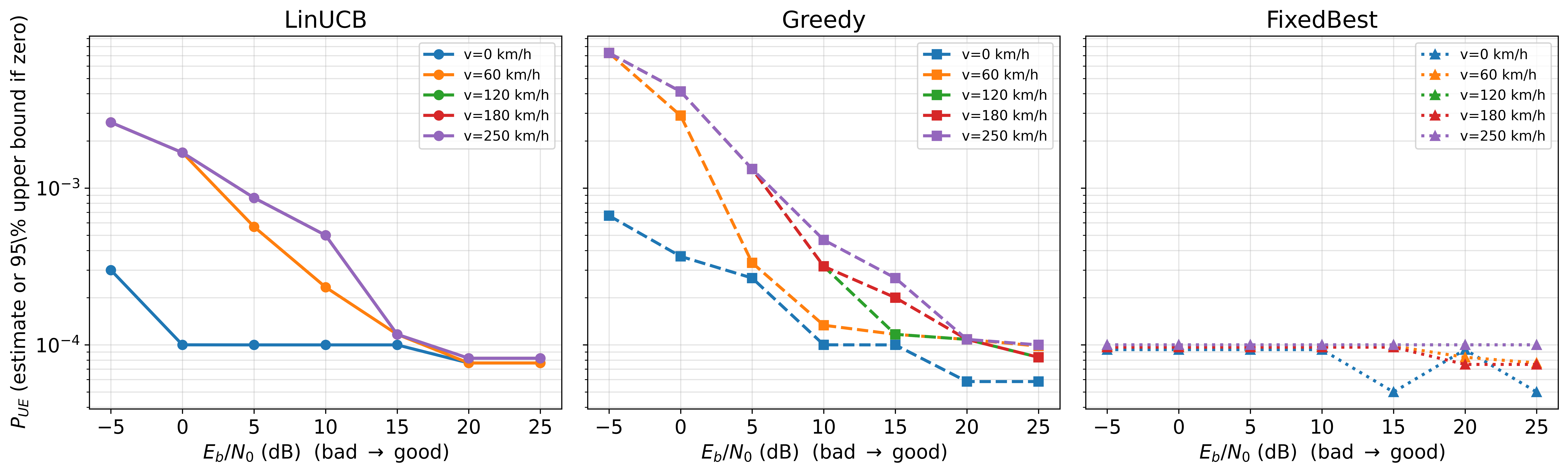}
    \caption{Rare-event evaluation of $P_{UE}$ vs. $E_b/N_0$ under different mobility regimes.}
    \label{fig:pue_vs_ebn0_all}
\end{figure*}

Fig.~\ref{fig:pue_vs_ebn0_all} reports rare-event estimates of $P_{UE}$ vs. $E_b/N_0$ for all policies and mobility regimes. Three regime-level observations emerge:
\begin{itemize}
\item {Low mobility ($v=0$~km/h):} all policies achieve low $P_{UE}$ as $E_b/N_0$ increases. LinUCB closely tracks FixedBest and improves over greedy in the low-to-mid $E_b/N_0$ region, indicating that uncertainty-aware exploration does not compromise integrity in quasi-stationary conditions.
\item {Moderate mobility ($v=60$ and $120$~km/h):} learning-assisted adaptation provides the clearest integrity benefit. In the low $E_b/N_0$ regime ($-5$ to $5$~dB), LinUCB achieves substantially lower $P_{UE}$ than greedy, consistent with its discounting and confidence-driven exploration. At higher $E_b/N_0$ (10–25~dB), LinUCB approaches FixedBest, indicating that online adaptation remains effective as UEs become rare.
\item {High mobility ($v=180$ and $250$~km/h):} the advantage of online adaptation diminishes. LinUCB and greedy exhibit similar $P_{UE}$, while FixedBest provides the most stable UE protection across $E_b/N_0$. This behavior is consistent with the training dynamics in Fig.~\ref{fig:running_pue_all}: when temporal correlation is weak, recent observations become less predictive, limiting the effectiveness of learning.
\end{itemize}

\begin{figure}[!htbp]
\centering \includegraphics[width=\linewidth]{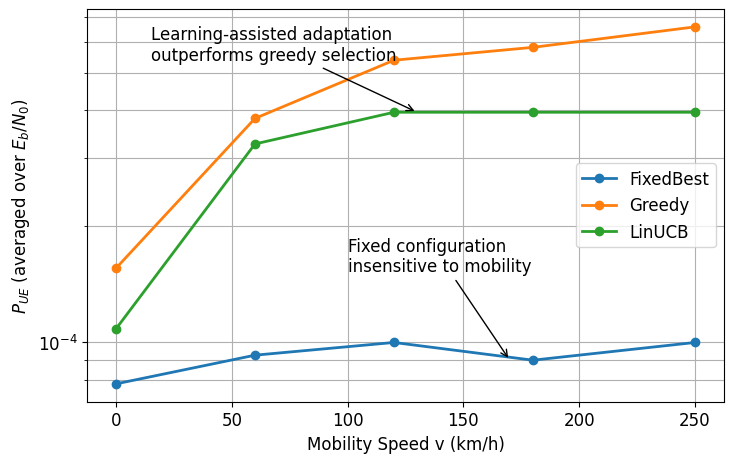} \caption{Rare-event evaluation of $P_{UE}$ vs. vehicle speed under averaged SNR values.} \label{fig:pue_vs_mob}
\end{figure}

To further isolate the effect of mobility, Fig.~\ref{fig:pue_vs_mob} summarizes $P_{UE}$ vs. speed using SNR-averaged values. The greedy policy exhibits the strongest sensitivity to speed, reflecting its myopic response to noisy short-term feedback. LinUCB consistently improves over greedy at low and moderate mobility, but the gap narrows at high mobility, where $P_{UE}$ saturates. In contrast, FixedBest remains largely insensitive to speed, with substantially smaller variation across the entire range. This confirms the intended trade-off: adaptation is most valuable when the channel is non-stationary but still predictable over the feedback horizon; when mobility breaks this predictability, conservative configuration becomes the safer integrity strategy.

\subsection{Interpreting Configuration Choices Across Mobility Regimes}
To explain the regime-dependent trends observed above, Table~\ref{tab:preferred_configs} summarizes the CRC–QC-LDPC configurations preferred by each policy across mobility levels. FixedBest converges to a single conservative pairing for mobility regimes beyond 60~km/h, explaining its stability under increasing speed. LinUCB, in contrast, adapts its preferred configuration as mobility increases, generally shifting toward stronger CRCs and lower code rates as the channel becomes less temporally correlated. This adaptive behavior is beneficial at moderate mobility, where contextual feedback retains predictive value, but yields diminishing returns at high mobility, where frequent configuration changes cannot reliably track fast channel evolution. These findings reinforce the core finidng that learning-assisted joint configuration can improve 5G-NR integrity under realistic mobility, but must be deployed with awareness of the mobility regime, since severe non-stationarity would favor conservative fixed configurations.

\begin{table*}[!htbp]
\caption{Preferred CRC-QC-LDPC Configurations Across Mobility Regimes}
\label{tab:preferred_configs}
\centering
\begin{tabular}{c|c|c}
\hline
\textbf{Speed (km/h)} &
\textbf{FixedBest Configuration} &
\textbf{LinUCB Preferred Configuration} \\
\hline
0   & CRC11 + $R=3/4$ ($k=432,n=576$) & CRC11 + $R=3/4$ ($k=432,n=576$) \\
60  & CRC16 + $R=3/4$ ($k=432,n=576$) & CRC24A + $R=1/2$ ($k=288,n=576$) \\
120 & CRC16 + $R=3/4$ ($k=432,n=576$) & CRC11 + $R=2/3$ ($k=384,n=576$) \\
180 & CRC16 + $R=3/4$ ($k=432,n=576$) & CRC11 + $R=1/2$ ($k=288,n=576$) \\
250 & CRC16 + $R=3/4$ ($k=432,n=576$) & CRC16 + $R=2/3$ ($k=384,n=576$) \\
\hline
\end{tabular}
\end{table*}

\section{Conclusion and Future Work}\label{conc}
In this study, we examined whether a 5G-NR V2X communication system can reduce silent (undetected) errors by adapting its physical-layer error-detection and error-correction settings {online} as vehicles move and channel conditions change. Using only standard-compliant CRC polynomials and QC-LDPC coding rates, we formulated joint configuration selection as a lightweight CB problem and evaluated a discounted LinUCB policy against a greedy online baseline and a conservative fixed configuration in a Sionna-based 5G-NR physical-layer simulation with mobility-induced time-correlated fading and non-stationary interference. Results show that learning-assisted adaptation is effective at low to moderate mobility, where recent feedback remains predictive: in this regime, LinUCB reduced the $P_{UE}$ by up to 50–70\% compared to greedy selection in the low-SNR range ($-5$ to 5~dB) and closely approached the best fixed configuration at higher $E_b/N_0$. These gains arise because the learning agent adapts toward stronger CRCs and lower coding rates, such as CRC24A with $R=1/2$ at 60~km/h and CRC11 with $R=2/3$ at 120~km/h. At high mobility ($\geq 180$~km/h), channel conditions decorrelate rapidly, the benefit of online learning diminishes, and a conservative fixed strategy consistently selecting CRC16 with $R=3/4$ provides the most stable protection across SNR. Overall, the results indicate that learning-assisted joint CRC–QC-LDPC configuration can improve physical-layer integrity in V2X systems when mobility-induced non-stationarity is moderate, but must be mobility-aware, as severe non-stationarity favors conservative fixed operation. Future work will extend this framework to multi-link and multi-vehicle scenarios, incorporate more realistic contention-based delayed feedback, and develop mobility-aware mechanisms that safely switch between adaptive and fixed configurations.

\section*{Acknowledgment}
This work was supported by the Research Fund of the Istanbul Technical University. Project Number: 47198. The authors thank Prof. Khaled Abdel-ghaffar from the Department of Electrical and Computer Engineering, University of California Davis, for his valuable comments regarding the encoding background.
\bibliographystyle{unsrt}
\bibliography{main}
\end{document}